\documentclass[11pt]{article}
\pdfoutput=1
\usepackage{jcappub,natbib}
\input{colordvi.tex}

\def\ga{\mathrel{\raise.3ex\hbox{$>$\kern-.75em\lower1ex\hbox{$\sim$}}}}
\def\la{\mathrel{\raise.3ex\hbox{$<$\kern-.75em\lower1ex\hbox{$\sim$}}}}

\title{Gravitational leptogenesis in Natural Inflation}

\author[a]{Alexandros Papageorgiou,}
\author[a]{Marco Peloso}

\affiliation[a]{School of Physics and Astronomy, and Minnesota Institute for Astrophysics, University of Minnesota, Minneapolis, 55455 (USA)}

\abstract{We compute the gravitational leptogenesis generated from the parity-violating gravitational waves sourced by an abelian gauge field coupled to a pseudo-scalar inflation. We show that, once the CMB bound on the tensor-to-scalar ratio is enforced, the lepton asymmetry produced by this mechanism during inflation is too small  to account for the observed baryon asymmetry of the universe, irrespectively of the inflaton potential, the strength of its coupling to the gauge field, and the details of reheating.}

\begin{document}

\begin{flushright}   UMN-TH 3634/17  \end{flushright}

\maketitle
\flushbottom

\section{Introduction}
\label{sec:intro} 

One of the open questions in cosmology is the origin  of the matter-antimatter asymmetry of the universe. While
Big Bang nucleosynthesis has traditionally provided the primary determination of the baryon asymmetry, since the WMAP measurements the Cosmic Microwave Background (CMB) gives the most precise value of the asymmetry \cite{Cyburt:2003fe}. The current Planck measurements give $\frac{n_B}{n_\gamma} = \left( 6.09 \pm 0.06 \right) \cdot 10^{-10}$ \cite{Ade:2015xua,Olive:2016xmw} for the ratio between the baryon and the photon number density. 

Various mechanisms of baryongenesis have been proposed in the literature. In this work we study a mechanism of leptogenesis proposed in \cite{Alexander:2004us} (see also \cite{Lyth:2005jf,Maleknejad:2014wsa,Kawai:2017kqt}), which makes use of the axial-gravitational anomaly \cite{AlvarezGaume:1983ig}. We study the embedding of this mechanism in the context of natural inflation \cite{Freese:1990rb}. In these models the inflaton is a pseudo-scalar field $\phi$ with a technically naturally flat potential (due to an an approximate shift symmetry). Reheating in natural inflation is typically dominated by the coupling $\phi F {\tilde F}$ of the pseudo-scalar inflaton to gauge fields  (the other dimension $5$ operator $\partial_\mu \phi {\bar \psi} \gamma^\mu \gamma^5 \psi$  that couples the inflaton to an axial fermionic current results in a decay that is helicity suppressed, see for instance the discussion  in \cite{Pajer:2013fsa}). The same coupling can lead to several interesting observational effects already during inflation, including large non-gaussianity \cite{Barnaby:2010vf}, running of the scalar perturbations \cite{Meerburg:2012id},  gravitational waves (GW)  \cite{Namba:2015gja} at CMB scales, as well as GW \cite{Cook:2011hg,Bartolo:2016ami} and primordial black holes at smaller scales \cite{Linde:2012bt,Garcia-Bellido:2017aan}. The GW produced by this mechanism are chiral \cite{Sorbo:2011rz}. This acts as a source for lepton asymmetry through the axial-gravitational anomaly \cite{AlvarezGaume:1983ig} that we quantify in the present work.~\footnote{This mechanism differs from the one considered in \cite{Giovannini:1997eg,Anber:2015yca,Kamada:2016eeb,Kamada:2016cnb,Cado:2016kdp,Jimenez:2017cdr}, in which a pseudo-scalar inflaton is coupled to the Standard Model hypercharge, which then sources the baryon asymmetry.}

The mechanism of gauge field amplification in natural inflation that is at the basis of the present study has strong analogy with that of gauge-flation \cite{Maleknejad:2011jw} and chromo-natural inflation \cite{Adshead:2012kp}. These models are characterized by an SU(2) gauge field, where the three gauge field components have a spatially-isotropic combination of vevs $A_i^a \propto \delta_a^i$ (where $a$ and $i$ are, respectively, spatial and $SU(2)$ indices).~\footnote{On the contrary, in the mechanisms studied here the gauge field is abelian, and it has no vev.}  Although the original versions of these models are ruled out \cite{Dimastrogiovanni:2012ew,Adshead:2013nka,Namba:2013kia}, new versions have been constructed, \cite{Maleknejad:2016qjz,Maleknejad:2016dci,Dimastrogiovanni:2016fuu,Obata:2016oym,Adshead:2017hnc,Caldwell:2017chz}. These models also result in a parity violating GW background, and, in particular, the works  \cite{Maleknejad:2016dci} and  \cite{Caldwell:2017chz} studied whether this can lead to a sufficiently large lepton asymmetry. While the estimates of \cite{Maleknejad:2016dci} indicate a very small production, a much greater asymmetry is produced with the inflaton potential studied in  \cite{Caldwell:2017chz}. These results provide the motivation for the present study. On one hand, we want to to understand whether this mechanism of baryogengesis can be successful also in the simpler context of natural inflation. On the other hand, we want to understand whether the signatures studied in the  works listed above are compatible with possible limits from lepton overproduction. 

The plan of this work is as follows. In Section \ref{sec:GW} we review the production of a parity violating GW background from gauge fields in natural inflation. In Section \ref{sec:igravilepto} we compute the baryon asymmetry generated during inflation by this mechanism. In Section \ref{sec:reheating} we show that there is a limit on how much this asymmetry can grow during reheating, even accounting for unconventional evolution of the inflaton, and of its decay products during this period. In Section \ref{sec:upper} we show that this mechanism cannot account for the baryon asymmetry of the universe. In Section \ref{sec:models}, for illustrative purposes,  we compute the asymmetry produced by this mechanism with some representative inflaton potentials.  In Section \ref{sec:icocnlusions} we present our conclusions.

\section{Sourced Gravitational waves in axion inflation} 
\label{sec:GW} 

We consider an axion inflaton coupled to a U(1) gauge field, with action 
\begin{equation}
\mathcal{S}= \int d^4 x \sqrt{-g} \left[\frac{M_p^2}{2}R -\frac{1}{2}\left(\partial\phi\right)^2-V\left(\phi\right)-\frac{1}{4}F^{\mu\nu} F_{\mu\nu}-\frac{1}{4f}\phi\tilde{F}^{\mu\nu}F_{\mu\nu}\right] \;, 
\end{equation}
where $\phi$ is the inflaton (assumed to be a pseudo-scalar), and $\tilde{F}= \frac{1}{2}\frac{\eta^{\mu\nu\alpha\beta}}{\sqrt{-g}}F_{\alpha\beta}$ is the dual of the gauge field strength tensor, with $\eta^{0123}=1$. Due to the $\phi F {\tilde F}$ coupling, the motion of the inflaton leads to an instability for one polarization of the gauge field, which during inflation is well described by \cite{Anber:2006xt}
\begin{equation}
A_+\left(\tau,k\right)\cong\frac{1}{\sqrt{2k}}\left(\frac{k}{2\xi a H}\right)^{1/4}e^{\pi\xi-2\sqrt{2\xi k/\left(a H\right)}} 
\;\;,\;\; \xi \equiv \frac{\dot{\phi}}{2 H f} \;, 
\label{Apsol}
\end{equation}
in the interval $\left(8\xi\right)^{-1} \lesssim k/\left(a H\right) \lesssim 2\xi $ of phase space that accounts for most of the power in the produced gauge fluctuations \cite{Barnaby:2011vw}. Note that the other polarization $A_-$ is not produced and therefore can be ignored.~\footnote{We are assuming $\xi > 0$. If instead $\xi < 0$ the other gauge field polarization is produced, resulting in a sourced gravitational wave field background with opposite chirality with respect to the one obtained here.}

The produced gauge quanta source scalar (curvature) and tensor (gravity waves, GW) perturbations \cite{Barnaby:2010vf,Barnaby:2011vw}. Due to the breaking of parity associated with the $A_+ $ production, which is ultimately related to the breaking of parity due to the motion of the pseudo-scalar inflaton, the left handed GW chirality is produced in a much greater amount than the right handed \cite{Sorbo:2011rz} one. The total GW power spectrum (vacuum $+$ sourced modes) is given by 
\begin{equation}
P_{GW} = P_{h,L}+P_{h,R} \simeq \frac{2H^2}{\pi^2 M_p^2}\left(\frac{k}{k_0}\right)^{n_T}\left[1+\frac{H^2}{M_p^2}f_{h,L}\left(\xi\right)e^{4\pi\xi}\right] \;\;, 
\label{PLR} 
\end{equation}
where we have disregarded the sourced right handed GW, and where 
\begin{equation}
f_{h,L}\cong \frac{4.3 \cdot 10^{-7}}{\xi^6}, \,\,\,\, \xi \gg 1 \;. 
\end{equation}
In eq. (\ref{PLR}), the pivot scale $k_0$ and the tensor tilt $n_T$ parametrize the vacuum GW, which are produced by the expansion of the universe in the standard way. The parity violating sourced GW background gives a nonvanishing Pontryagin density. Using the line element $ds^2 = a^2(\tau) \left[ -d\tau^2+(\delta_{ij}+\gamma_{ij})dx^i dx^j \right]$, one finds 
\begin{eqnarray} 
R \tilde{R} &\equiv& R_{\alpha\beta\rho\sigma}R_{\gamma\delta\mu\nu}\frac{\epsilon^{\alpha\beta\gamma\delta}}{\sqrt{-g}}g^{\mu\rho}g^{\nu\sigma} = - \frac{2}{a^4}\epsilon^{ijk} \, \big( \partial_\tau^2 \gamma_{jl} \, \partial_i\partial_\tau \gamma_{lk}-\partial_m \partial_\tau \gamma_{jl} \partial_i\partial_m\gamma_{lk}+\partial_l \partial_\tau \gamma_{jm} \partial_m\partial_i\gamma_{kl}\big)
 \,.  \nonumber\\ 
\label{RRtilde}
\end{eqnarray} 
As we show in  Appendix \ref{app:graviL}, the expectation value for this quantity is different from zero if the left-handed and the right-handed GW are produced in different amount.  This acts as a source for the lepton asymmetry, as we discuss in the next section.

\section{Gravi-leptogenesis during inflation} 
\label{sec:igravilepto} 

The gravitational anomaly leads to a non-conservation of the lepton number \cite{AlvarezGaume:1983ig}
\begin{equation}
\nabla_\mu J^\mu_L=\frac{1}{24} \times \frac{N_{R-L}}{16\pi^2}R\tilde{R} \;, 
\label{gravi-anomaly} 
\end{equation}
where $N_{R-L}$ is the number of right-handed minus left-handed particles. In the Standard Model  $N_{R-L}= -3$, due to the absence of right-handed neutrinos. We can assume that right-handed neutrinos are present, but that the Hubble rate during inflation is smaller than their mass. Therefore we will set  $N_{R-L} = - 3$ in this work.~\footnote{As we show in Appendix \ref{app:correl}, our result for the produced asymmetry is dominated by modes that are of the size of the horizon, $k \simeq H$, at the end of inflation. As long as the Hubble rate during inflation is smaller than the right-handed neutrino masses (which is an assumption compatible with the measured light neutrinos $\Delta m^2$ and the see-saw mechanism), we can use the relation 
(\ref{gravi-anomaly}) with $N_{R-L}=-3$ in our computations.} 

Starting from this expression, the computation  presented in Appendix \ref{app:graviL} leads to the lepton number 
\begin{equation}
n_L =  \frac{N_{R-L}}{192 \, \pi^2 a^3} \int d \tau \int d \ln k \; 
\sum_{\lambda=\pm} \lambda  \, \left[ k \,  P_\lambda^{(2,1)} \left( \tau ,\, k \right)  -  k^3   P_\lambda^{(1,0)}\left( \tau ,\, k \right) \right] \;, 
\label{nL}
\end{equation}
where the sum is performed over the two GW helicities, and where we have defined the quantities
\begin{equation}
\left\langle \left( \partial_\tau \right)^m h_\lambda \left( \tau,\vec{k} \right) \;  \left( \partial_\tau \right)^n h_\sigma \left( \tau,\vec{k}' \right) 
\right\rangle \equiv \delta_{\lambda \sigma}  \delta^{(3)} \left( \vec{k} + \vec{k}' \right) \; \frac{2 \pi^2}{k^3} \, P_\lambda^{(m,n)} \left( \tau ,\, k \right) \;. 
\label{P-mn} 
\end{equation} 
We note that $P_\lambda^{(0,0)}$ is the power spectrum of the helicity $\lambda$, while the two upper indices $m$ and $n$ indicate how many time derivatives are acting on the two mode functions. 

The lepton number is partially converted into baryon number by sphalerons. For the standard model, one finds the baryon abundance \cite{Kuzmin:1985mm,Khlebnikov:1988sr} 
\begin{equation}
Y_B = -\frac{28}{79} \, Y_L  = - \frac{28}{79} \, \frac{n_L}{s} \,, 
\end{equation}
where $s$ is the entropy density of the thermal bath formed at reheating. The abundance is constant, from the end of reheating onwards, as both the numerator and denominator scale as inverse volume. From the measured value $ \frac{n_B}{n_\gamma} = \left( 6.09 \pm 0.06 \right) \cdot 10^{-10}$ \cite{Ade:2015xua,Olive:2016xmw} (where $n_\gamma$ is the photon number density), and from the relation $s \simeq 7.04 n_\gamma$, we see that a successful mechanism of leptogenesis must produce 
\begin{equation}
Y_{L,{\rm needed}} \simeq -2.44 \cdot 10^{-10} \,. 
\end{equation} 
Combining the above expressions, we can express the ratio of the lepton asymmetry produced by the gravitational anomaly over that required to explain the observations, 
\begin{equation}
\frac{Y_L}{Y_{L,{\rm needed}}} = 
8.8 \cdot 10^5 \,  \left( \frac{H}{M_p} \right)^{3/2} \, 
 \frac{1}{a^3 H^3} \, \int d \tau \int d \ln k \sum_{\lambda = \pm} \; 
 \lambda  \, \left[ k \,  P_\lambda^{(2,1)} \left( \tau ,\, k \right)  -  k^3   P_\lambda^{(1,0)}\left( \tau ,\, k \right) \right] \;. 
\label{YYL-par1}
\end{equation}
We note that, in writing this expression, we are implicitly dividing the lepton number density by the entropy density at all times (including during inflation). Strictly speaking, the entropy density should be introduced only after thermalization has completed. During reheating, the total energy of the universe is converted into a thermal bath, of energy density $\rho$ and of entropy density $s \propto \rho^{3/4}$. Only at the formation of this thermal bath we can properly define the lepton abundance $Y_L \equiv \frac{n_L}{s}$. However, given that $s \propto \rho^{3/4}$ at the end of reheating, studying the evolution of $\frac{n_L}{\rho^{3/4}}$ during inflation and reheating allows to study the time evolution of the quantity that eventually becomes $Y_L$ once reheating has completed. With this understanding, we define $Y_L \propto n_L / \left( \rho_\phi + \rho_A \right)^{3/4}$ at all times. 

The expression (\ref{YYL-par1}) is evaluated in  Appendix \ref{app:correl}, where we obtain 
\begin{equation}
\frac{Y_L}{Y_{L,{\rm needed}}} \Bigg\vert_{\rm end \; inflation} \simeq  
0.051 \, \, \left( \frac{H_{\rm end}}{M_p} \right)^{11/2} \, \int_{0}^{1} d a \, a^2  \, \left( \frac{H}{H_{\rm end}} \right)^7 \;  \frac{e^{4\pi\xi}}{\xi^{6.52}}  \;\;\;,\;\;\; 3 \leq \xi \leq 7 \,, 
\label{YYL-par2}
\end{equation}
in the interval relevant for our discussion (see below), and where we have normalized the scale factor to one at the end of inflation.~ 

In typical models, $\xi$ grows during inflation (in the regime of weak backreaction of the produced gauge fields on the background evolution, $\xi \propto \frac{\dot{\phi}}{H} \propto \sqrt{\epsilon}$, where $\epsilon \equiv \frac{M_p^2}{2} \left( \frac{1}{V} \, \frac{d V}{d \phi} \right)^2 $ is the standard slow roll parameter).   Therefore, due to the exponential dependence on $\xi$, and due to the phase space giving rise to the $a^2$ factor in eq. (\ref{YYL-par2}),  the amount of asymmetry produced during inflation is dominated by the final stages. Disregarding the variation of $H$ and $\xi$ in this final stage~\footnote{This can be done as long as they vary adiabatically, namely as long as $\epsilon_H \equiv - \frac{\dot{H}}{H^2}$ and $\epsilon_\xi \equiv \frac{\dot{\xi}}{H \, \xi}$ are $\ll 1$; the examples discussed in Section \ref{sec:models} have $\epsilon_H \ll 1$ until the end of inflation (when $\epsilon_H$ approaches one), and $\epsilon_\xi \ll 1$ all throughout inflation. Treating $H_{\rm end}$ and $\xi_{\rm end}$ as constant maximizes the production. We will see below that even in this case this mechanism produces an insufficient amount of asymmetry.}  results in 
\begin{equation}
\frac{Y_L}{Y_{L,{\rm needed}}}  \Bigg\vert_{\rm end \; inflation} \simeq  
0.017  \, \left( \frac{H_{\rm end}}{M_p} \right)^{11/2} \,  \frac{e^{4\pi\xi_{\rm end}}}{\xi_{\rm end}^{6.52}}  \;\;\;,\;\;\; 3 \leq \xi \leq 7 \;. 
\label{YYL-constant-xi-H}
\end{equation}
%

\section{Evolution at reheating} 
\label{sec:reheating} 

In this section we discuss the evolution during reheating of the lepton asymmetry generated during inflation. We do not discuss the possible generation of additional asymmetry at reheating that could take place for instance if the gauge field is further amplified by coherent oscillations of the inflation about the minimum of its potential (preheating). A full study of this effect can be performed through lattice simulations, along the lines of \cite{Adshead:2015pva}. 

Reheating is a continuous process that accounts for the decay of the inflaton field. It  starts at the end of inflation, and it ends at the formation of a dominant thermal bath of temperature $T_{\rm reh}$ and of equation of state $w \equiv \frac{p}{\rho} = 1/3$ (where $p$ is pressure and $\rho$ energy density). The total equation of state $w \left( t \right)$ varies in a model dependent way during reheating. For instance a massive inflaton field that is still dominant and that oscillates about the minimum of the potential leads to an equation of state that oscillates with average $w=0$, so that perturbative reheating leads $w$ to adiabatically change from $w=0$ to $w=1/3$. Quick preheating typically leads to an equation of state that is intermediate between these values already after the first few inflaton oscillations \cite{Podolsky:2005bw}. On the contrary, a rapid decrease of the inflaton potential after inflation can lead to a phase of kinetion, characterized by $w=1$. 

Many phenomenological studies of reheating introduce two parameters: (i) a constant (average) equation of state $w_X$ during reheating, and (ii) the reheating temperature  $T_{\rm reh}$. We now discuss the evolution of the lepton asymmetry generated during inflation under this  parametrization. 

We use as a ``time variable'' the number of e-folds $N$.  To respect the conventions chosen for inflation, we set at all times 
\begin{equation}
N = - \ln \frac{a}{a_{\rm end \, inflation}} \;\;, 
\end{equation}
where $a$ is the scale factor. Therefore $N=0$ at the end of inflation, and $N<0$ afterwards. Any inflationary evolution leads to three parameters of interest for our discussion (all of them evaluated at the end of inflation): (i) the net lepton number density $n_L$, (ii) the energy density in the inflaton  $\rho_\phi$, and (iii) the energy density in the gauge field $\rho_A$. We denote the ratio between these two energy densities as 
\begin{equation}
r_A \equiv \frac{\rho_A}{\rho_\phi} \,. 
\end{equation} 

Under the above assumptions, the lepton number density is diluted by the volume of the universe as $n_L \propto {\rm e}^{3 N}$. The energy density in the inflaton and in its decay products instead evolves as $\rho_\phi \propto {\rm e}^{3 \left( 1 + w_X \right) N}$.~\footnote{We stress that, during inflation, $\rho_\phi$ denotes the energy density of the inflaton field. During reheating it denotes the sum of the energy density of the inflaton and of its decay products, and $w_X$ denotes the average equation of state of this mixture during reheating.}  The gauge field is a massless species, whose energy density evolves as $\rho_A \propto {\rm e}^{4 N}$. As a consequence, during reheating $(N < 0$), the lepton asymmetry and the fraction of energy density in the gauge field evolve as 
\begin{equation}
\left\{ \begin{array}{l} 
Y_L \left( N \right) =  Y_L \left( 0 \right)  \,  \, {\rm e}^{-\frac{3}{4} N \left( 3 w_X - 1 \right)} 
\left[ \frac{1+r_A \left( 0 \right)}{1+r_A \left( 0 \right) \, {\rm e}^{-N \left( 3 w_X - 1 \right)} } \right]^{3/4} 
 \;\;, \\ \\ 
r_A \left( N \right) = r_A \left( 0 \right) \, {\rm e}^{-N \left( 3 w_X - 1 \right)} \;. 
\end{array} \right. 
\label{YL-Ra-reheating}
\end{equation} 

We are interested in the maximum possible asymmetry produced by this mechanism, so here we consider the possibility that the asymmetry grows during reheating. This is possible if  $w_X > \frac{1}{3}$. We see from  (\ref{YL-Ra-reheating}) that this implies that also  the ratio $r_A$ grows during reheating.  The two relations (\ref{YL-Ra-reheating}) can be combined into 
\begin{equation}
Y_L \left( N \right) =  Y_L \left( 0 \right) \times \left[ \frac{r_A \left( N \right)}{r_A \left( 0 \right)} \right]^{3/4} 
\left[ \frac{1 + r_A \left( 0 \right)}{1+r_A \left( N \right)} \right]^{3/4} \;. 
\label{YL-reheating} 
\end{equation} 

This quantity can grow only until $r_A \left( N \right) \simeq 1$, after which it assumes the constant  value 
$Y_L \left( N \right) \simeq \frac{Y_L \left( 0 \right)}{r_A^{3/4} \left( 0 \right)}$. As long as $r_A \left( N \right) < 1$, the  ratio $\frac{r_A \left( N \right) \left[ 1 + r_A \left( 0 \right) \right]}{1+r_A \left( N \right)}$ is always smaller than one. Therefore from eq. (\ref{YL-reheating})  we obtain an upper bound on the asymmetry 
\begin{equation}
Y_{L,{\rm max }} = \frac{Y_{L,{\rm end \, inflation}}}{r_{A,{\rm end \, inflation}}^{3/4}} \;. 
\label{YL-max}
\end{equation} 
This is a general upper bound, that holds irrespectively of the detailed evolution of inflation (namely, irrespectively of $w_X$ and of $T_{\rm reh}$).~\footnote{We note that this upper bound on the enhancement of the asymmetry appears to be violated in  \cite{Maleknejad:2016dci} (at least, for the numerical values considered at page 13 of that work).  Ref.   \cite{Maleknejad:2016dci}  obtains the observed amount of the asymmetry provided that the asymmetry  grows by a factor $\sim 10^9 \,\, - \,\, 10^{13}$ during reheating. On the other hand, the mechanism considered in that work also results in a gauge field energy density that satisfies $r_A \left( 0 \right) \simeq 10^{-4}$ at the end of  inflation.}  It only assumes that (i) leptogenesis is effective only during inflation (namely, leptons and anti-leptons are produced in the same amount during reheating), and (ii) the gauge field produced by the $\phi F {\tilde F}$ coupling has an equation of state $w=1/3$ all throughout reheating.~\footnote{In principle, one could imagine a situation in which the gauge field (once given a mass that was negligible during inflation) could decay into a species with $w > 1/3$. This, however, requires a series of ad hoc assumptions, and therefore we disregard this possibility in the present study.}  

Using the mode function (\ref{Apsol}), we find during inflation \cite{Barnaby:2011vw} 
\begin{equation}
\rho_A \simeq 1.4 \cdot 10^{-4} \, \frac{H^4}{\xi^3} \, {\rm e}^{2 \pi \xi} \;\;. 
\end{equation}
Using this, plus $\rho_\phi = 3 H^2 M_p^2 - \rho_A$, we find 
\begin{equation}
r_{A,{\rm end \, inflation}} \simeq \frac{4.7 \cdot 10^{-5} \, \left( \frac{H_{\rm end}}{M_p} \right)^2 \, \frac{{\rm e}^{2 \pi \xi_{\rm end}}}{\xi_{\rm end}^3}}{1 - 4.7 \cdot 10^{-5} \, \left( \frac{H_{\rm end}}{M_p} \right)^2 \, \frac{{\rm e}^{2 \pi \xi_{\rm end}}}{\xi_{\rm end}^3}} \, . 
\label{rA-end}
\end{equation}

\section{Upper bound on the produced asymmetry  }
\label{sec:upper} 

In Section \ref{sec:igravilepto} we found the relation 
\begin{equation}
\frac{Y_L}{Y_{L,{\rm needed}}}  \Bigg\vert_{\rm end \; inflation} \leq 
0.017  \, \left( \frac{H_{\rm end}}{M_p} \right)^{11/2} \,  \frac{e^{4\pi\xi_{\rm end}}}{\xi_{\rm end}^{6.52}}  \;\;\;,\;\;\; 3 \leq \xi_{\rm end} \leq 7 \;. 
\label{Y-end}
\end{equation}
for the lepton asymmetry at the end of inflation. In Section \ref{sec:reheating} we noted that the asymmetry produced during inflation can grow at reheating, but that it cannot exceed the value 
\begin{equation}
\frac{Y_{L,{\rm max}}}{Y_{L,{\rm needed}}} \leq 
30 \, \left( \frac{H_{\rm end}}{M_p} \right)^4 \, \frac{{\rm e}^{\frac{5}{2} \pi \xi}}{\xi^{4.27}} \, 
\left( 1 -  4.7 \cdot 10^{-5} \, \left( \frac{H_{\rm end}}{M_p} \right)^2 \, \frac{{\rm e}^{2 \pi \xi_{\rm end}}}{\xi_{\rm end}^3} \right)^{3/4} \,,  
\label{YL-YLneeded-final}
\end{equation} 
which is obtained by combining eqs. (\ref{YL-max}), (\ref{rA-end}), and (\ref{Y-end}).

\begin{figure}[ht!]
\centerline{
\includegraphics[width=0.5\textwidth,angle=0]{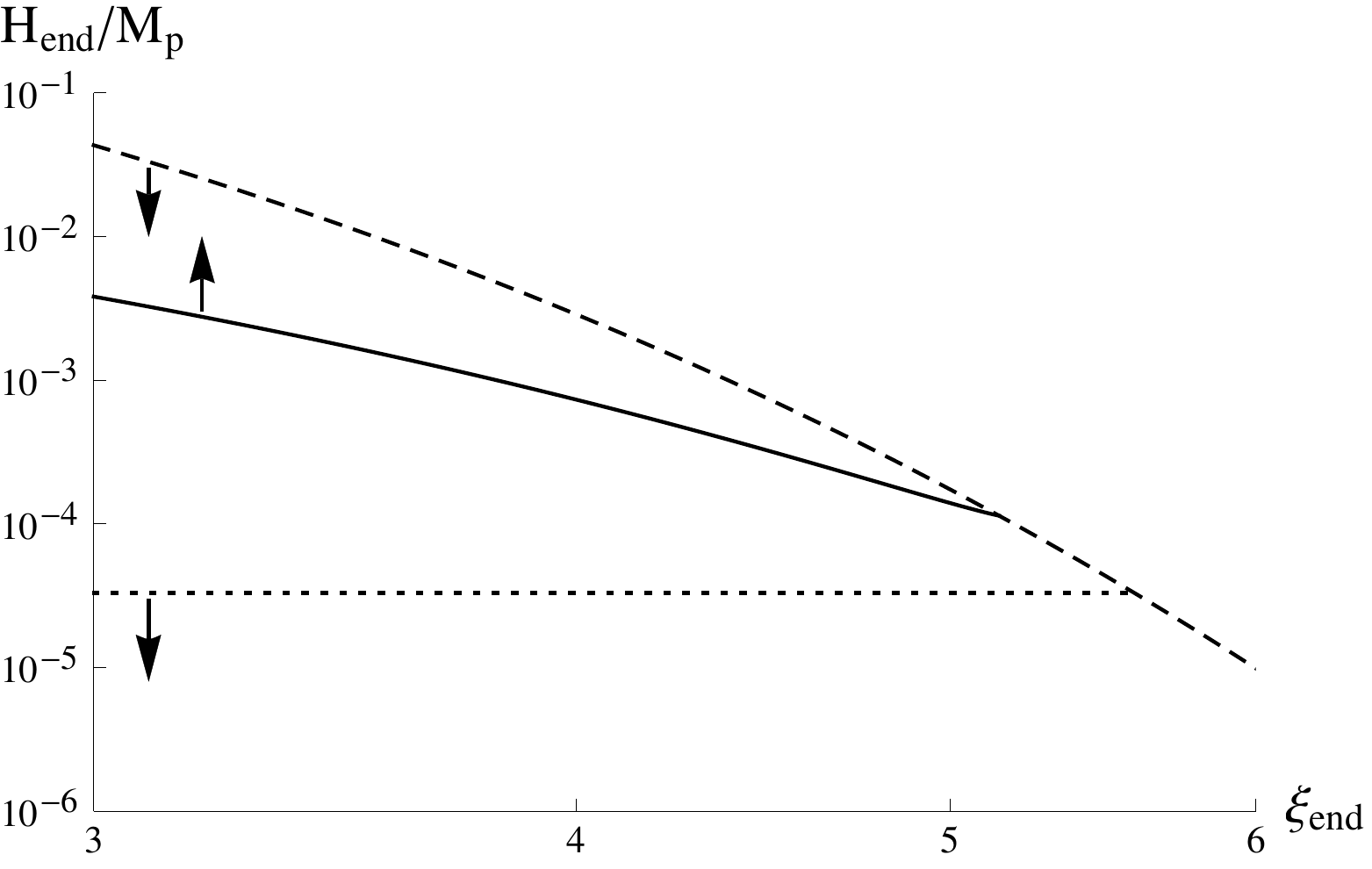}
}
\caption{Bounds on the value of the Hubble rate $H_{\rm end}$ at the end of inflation, for different values of the parameter $\xi_{\rm end}$ at the end of inflation. The dashed line is an upper bound obtained from eq. (\ref{rA-end}), resulting from the requirement that the energy density in the produced gauge fields is smaller than that of the inflaton. The solid line is a lower bound, obtained from eq. (\ref{YL-YLneeded-final}), resulting from the requirement that enough lepton asymmetry is produced. The dotted line is an upper bound, obtained from eq. (\ref{H-r}), resulting from the requirement that the tensor-to-scalar ratio $r < 0.1$. This assumes a constant $H$ during inflation. Accounting for the decrease of $H$ in typical inflationary model results in an even stronger bound. No values of $\xi_{\rm end}$ and $H_{\rm end}$ exist that satisfy all these bounds.}
\label{fig:analytic}
\end{figure}

For the leptogenesis mechanism to be successful we need to require that the r.h.s. of (\ref{YL-YLneeded-final}) is $\geq 1$. 
This translates into a lower limit on $H_{\rm end}$, that we show in the solid curve of Figure \ref{fig:analytic}.  At the same time, we must require that eq. (\ref{rA-end}) is $\leq 1$. This translates into an upper limit on $H_{\rm end}$, that we show in the dashed line of Figure \ref{fig:analytic}. Only the area in the $\left\{ \xi_{\rm end} - H_{\rm end} \right\}$ plane between these two curves can lead to a successful mechanism. In particular, this requires $\xi_{\rm end} \la 5.1$ and $H_{\rm end} \ga 10^{-4} \, M_p$. This limit on $H$ is however inconsistent with the CMB. At CMB scales the sourced scalar and tensor modes must be much smaller than the vacuum ones \cite{Barnaby:2010vf}, so that the standard slow roll relations can be used: 
\begin{equation}
r = 16 \epsilon \;\;,\;\; P_\zeta \simeq \frac{H^2}{8 \pi^2 \epsilon M_p^2} \simeq 2.2  \cdot 10^{-9} \;\; \Rightarrow \;\; 
\frac{H}{M_p} \simeq 3.3 \cdot 10^{-5} \, \left( \frac{r}{0.1} \right)^{1/2} \;, 
\end{equation}
where the primordial perturbations have been normalized to the value measured in \cite{Ade:2015lrj}, and $r$ is the tensor-to-scalar ratio. The quantity $H$ in this relation is the value assumed by the Hubble rate at the moment of CMB emission. The Hubble rate decreases during inflation, and therefore 
\begin{equation}
H_{\rm end}  \la 3.3 \cdot 10^{-5} \, \left( \frac{r}{0.1} \right)^{1/2} \;, 
\label{H-r} 
\end{equation}
This results in an upper bound on $H_{\rm end}$, that is shown as a dotted line in Figure  \ref{fig:analytic} (where $r < 0.1$ is taken 
\cite{Ade:2015xua}). This upper bound is incompatible with the region between the solid and the dashed curve. 

We conclude that, irrespectively of the inflaton potential, and of the reheating after inflation, this mechanism cannot explain the observed baryon asymmetry of the universe.

\section{Models  }
\label{sec:models} 

In this section we evaluate the asymmetry produced during inflation using specific inflationary potentials. We evolve the background equations consistently accounting for the backreaction of the produced gauge quanta in the evolution of the inflaton and the scale factor  (see for instance Section 2.2 of \cite{Barnaby:2011vw}), and we numerically integrate eq.  (\ref{YYL-par2}) for the asymmetry. Doing so, we consistently account for the time evolution of $H$ and $\xi$ (eq. (\ref{YYL-par2}) requires that $\xi$ is adiabatically evolving, $\epsilon_\xi \equiv  \frac{\dot{\xi}}{H \, \xi} \ll 1$; we have verified that the evolutions discussed in this section satisfy this). 

We choose potentials encountered in models of axion inflation. Specifically, we choose a quadratic potential (which can for instance be understood as the limit of the potential close to the minimum, which can be a good approximation for all observable inflation in the case of large axion decay constant), a linear potential (modified to be quadratic in the minimum; this is a typical potential for monodromy \cite{Silverstein:2008sg}) and a top hill potential (which can be obtained for instance for trajectories originating from saddle points in aligned axion inflation \cite{Peloso:2015dsa}). In all cases the scale of the potential is normalized to produce the correct amplitude of the scalar perturbations, for modes produced $60$ e-folds before the end of inflation. The coupling $\frac{1}{f}$ of the inflaton to the gauge field  is instead chosen to be as large as allowed by the Primordial Black Hole limit on the amount of scalar perturbations sourced by the gauge field, namely $\xi = 5$ at $8$ e-folds before the end of inflation \cite{Linde:2012bt,Garcia-Bellido:2016dkw}

\begin{table}
\centering
  \begin{tabular}{ | l | c | c | c | c | }
    \hline
    Potentials & $\xi_{\rm CMB}$  & $\xi_{\rm end}$ & $H_{\rm end}/M_p$ & $Y_{L,{\rm max}}/Y_{L,{\rm needed}}$ \\ \hline
    $V \left( \phi \right)=\frac{m^2}{2} \phi^2$ & 1.54 & 6.60 & $1.5 \times 10^{-6}$ & $2\times 10^{-3}$  \\ \hline
    $V \left( \phi \right)=m^2 M_p^2 \left( \sqrt{1-\frac{\phi^2}{M_p^2}}-1 \right)$ & 1.40  & 6.42 & $2.6 \times 10^{-6} $ & $5\times 10^{-3}$ \\ \hline
    $V \left( \phi \right)=m^2 M_p^2 \left( 1-\frac{\phi^2}{2 n} \right)$ & 1.41 & 6.37 & $3.7 \times 10^{-6}$ & $8\times 10^{-3}$ \\
    \hline
  \end{tabular}
  \caption{Values of the parameter $\xi$, of the Hubble rate, and of the maximum possible asymmetry obtained with some inflaton potentials. In all cases considered, the asymmetry is too small to explain the observations, in agreement with the general conclusions of the previous section.} 
    \label{table:results}
\end{table}

All the models considered here produce a too small amount of asymmetry, in agreement with the general result obtained in the previous section. Among the cases shown, the greatest result is obtained for the hill top potential, due to the fact that $H$ decreases less during inflation, and $H_{\rm end}$ assumes the greatest value. For the three models listed in Table \ref{table:results} we found, respectively, $r_A \left( 0 \right) = 0.37 ,\, 0.40 ,\, {\rm and } \, 0.59$. 
This implies that the asymmetry can increase only be an order one factor during reheating.

\section{Conclusions}
\label{sec:icocnlusions} 

In this work we studied whether the gravitational anomaly  \cite{AlvarezGaume:1983ig} can produce the observed baryon asymmetry of the universe in models of natural inflation. We considered a U(1) gauge field amplified by an axion inflaton through a typical $\phi F {\tilde F}$ coupling. The amplified gauge field sources parity-violating GW, which produce a lepton asymmetry through the gravitational anomaly. We computed the asymmetry produced at the end of inflation, also accounting for the possibility that the asymmetry increases during (unconventional) reheating. We found that, irrespectively of the inflaton potential, the strength of its coupling to the gauge field, and the details of reheating, this mechanism cannot account for the observed baryon asymmetry of the universe.  We conclude that a more conventional mechanisms for the asymmetry must be sought, and that the several phenomenological signatures that can be obtained from the gauge field amplification do not suffer from baryon overproduction. 

This result contrasts with what found in \cite{Caldwell:2017chz} for a specific version of chromo-natural inflation (namely, for a specific inflaton potential which is compatible with observations). While in the cases that we have studied the gravitational waves are produced by a $\delta A + \delta A \rightarrow \delta g$ interaction, in chromo-natural inflation they are produced by a quadratic $\delta A \delta g$ mixing (present due to the fact that vector fields have a vev in this model). Requiring a successful gravitational leptogenesis in  the model of  \cite{Caldwell:2017chz} is very interesting phenomenologically, since it implies that the sourced GW are at an observable level.  While tensor nongaussianity within this class of models has been studied in \cite{Agrawal:2017awz,Thorne:2017jft}, a full study of nonlinearities is still in order to confirm their viability. 

We conclude by pointing out one limitation of our analysis. In the mechanism we have studied, the production of the asymmetry is dominated by modes that leave the horizon at the end of inflation. We have studied how the asymmetry produced during inflation can evolve during reheating, showing that in this mechanism there is an upper bound on how much the asymmetry can grow during this stage, irrespectively of the inflaton potential and of the details of reheating. We did not study the possibility that the gauge field produced during inflation is further amplified at preheating, with a possible increase of the sourced GW and lepton asymmetry. A full computation of this effect will presumably require a numerical study, for instance along the lines of \cite{Adshead:2015pva}.

\vskip.25cm
\section*{Acknowledgements} 

We thank Robert Caldwell for useful discussions.  The work of A.P. was partially supported by a summer grant from the graduate program of the School of Astronomy and Physics of the University of Minnesota. The work of M.P. is partially supported from the DOE grant DE-SC0011842  at the University of Minnesota.

\vskip.25cm

\appendix

\section{Computation of lepton asymmetry from $R {\tilde R}$ }
\label{app:graviL}
We present here the calculation for the lepton number density $n_L$ starting from the equation  
(\ref{gravi-anomaly}) for the gravitational anomaly. We use the expression  (\ref{RRtilde})  for the Pontryagin density in terms of the tensor fluctuations of the metric, and take the expectation value 
\begin{equation}
\partial_\tau\left(a^3 n_L\right) = -\frac{N_{R-L}}{8\pi^2 (24)} \, \left\langle \epsilon^{ijk}\big( \partial_\tau^2 \gamma_{jl} \, \partial_i\partial_\tau \gamma_{lk}-\partial_m \partial_\tau \gamma_{jl} \partial_i\partial_m\gamma_{lk}+\partial_l \partial_\tau \gamma_{jm} \partial_m\partial_i\gamma_{kl}\big) \right\rangle \;. 
\end{equation}
To evaluate the expectation value we decompose the tensor mode as 
\begin{equation}
\gamma_{ij}(\tau,\vec{x})=\int \frac{d^3 k}{(2\pi)^{3/2}}e^{i\vec{k}\vec{x}}\,\,\,\sum_{\lambda=\pm}\,\,\, \Pi^*_{ij,\lambda}\Big(\hat{k}\Big)h_\lambda\Big(\tau,\vec{k}\Big) \;, 
\end{equation}
where the polarization operators $\Pi_{ij,\lambda} \left( {\hat k} \right) = \epsilon_{i,\lambda}^* \left( {\hat k} \right) \epsilon_{j,\lambda}^* \left( {\hat k} \right) $  are transverse and traceless.~\footnote{The quantities $\vec{\epsilon}_{\lambda} \left( {\hat k} \right) $ are polarization operators of masses spin one fields, and they satisfy 
$\vec{k} \cdot \vec{\epsilon}_\lambda \left( {\hat k} \right)= 0 ,\, \vec{k} \times \vec{\epsilon}_\lambda \left( {\hat k} \right) = - i \lambda k \vec{\epsilon}_\lambda \left( {\hat k } \right) ,\, \vec{\epsilon}_\lambda \left( - {\hat k } \right) =  \vec{\epsilon}_\lambda \left(  {\hat k } \right)^* ,\,  \vec{\epsilon}_\lambda \left( {\hat k } \right)^* \cdot  \vec{\epsilon}_{\lambda'} \left( {\hat k } \right) = \delta_{\lambda \lambda'} $.}   This gives 
\begin{eqnarray}
\partial_\tau\left(a^3 n_L\right) &=& -\frac{N_{R-L}}{8\pi^2 (24)}  \int\frac{d^3k}{(2\pi)^3} \, \frac{2 \pi^2}{k^3} \, \epsilon^{ijk} \sum_{\lambda = \pm} \Bigg[ -ik_i\Pi^*_{jl,\lambda}\Big(\hat{k}\Big) \, \Big(\tau,k\Big)\Pi^*_{lk,\lambda}\Big(-\hat{k}\Big) \Big(\tau,k\Big) \, P_\lambda^{(2,1)} \left( \tau ,\, k \right)  \nonumber\\ 
&&  +ik_m k_i k_m \Pi^*_{jl,\lambda}\Big(\hat{k}\Big) \, \Pi^*_{lk,\lambda}\Big(-\hat{k}\Big) \,  P_\lambda^{(1,0)} \left( \tau ,\, k \right)  -ik_l k_m k_i \Pi^*_{jm,\lambda}\Big(\hat{k}\Big) \, \Pi^*_{kl,\lambda}\Big(-\hat{k}\Big) \, P_\lambda^{(1,0)} \left( \tau ,\, k \right)  \Bigg] \,, \nonumber\\ 
\end{eqnarray}
where the quantities $P^{(m,n)}$ are given in eq. (\ref{P-mn}). The last term in the sum vanishes due to transversality. In the other two terms we use 
\begin{eqnarray}
&& \epsilon^{ijk}k_i\Pi^*_{jl,\lambda}\Big(\hat{k}\Big)=\epsilon^{kij}k_i\epsilon_{j,\lambda}\Big(\hat{k}\Big)\epsilon_{l,\lambda}\Big(\hat{k}\Big) =-i\lambda k \epsilon_{k,\lambda}(\hat{k})\epsilon_{l,\lambda}(\hat{k})=-i\lambda k \Pi^*_{kl,\lambda}\Big(\hat{k}\Big) \;, 
\end{eqnarray} 
and therefore (where $f_\lambda$ is a generic function of $\lambda$) 
\begin{equation}
\epsilon^{ijk}k_i\Pi^*_{jl,\lambda}\Big(\hat{k}\Big) \Pi^*_{lk,\lambda} \Big(-\hat{k}\Big) f_\lambda = 
-i\lambda k \Pi^*_{kl,\lambda}\Big(\hat{k}\Big) \Pi_{lk,\lambda} \Big(\hat{k}\Big) f_\lambda = - i \lambda k \, f_\lambda \,. 
\end{equation} 
This gives 
\begin{eqnarray}
\partial_\tau\left(a^3 n_L\right) &=& -\frac{N_{R-L}}{8\pi^2 (24)}  \int\frac{d^3k}{(2\pi)^3}  \, \frac{2 \pi^2}{k^3}  
\; \sum_{\lambda=\pm} \left( - i \lambda k \right) \left[ -i   \, P_\lambda^{(2,1)} \left( \tau ,\, k \right)  + i \, k^2  \,  P_\lambda^{(1,0)} \left( \tau ,\, k \right)  \right] \;, \nonumber\\ 
\end{eqnarray}
from which the expression (\ref{nL}) of the text is obtained.

\section{Calculation of the correlation functions}
\label{app:correl}

In this appendix we evaluate the two quantities $P_\lambda^{(2,1)} \left( \tau ,\, k \right) $ and $P_\lambda^{(1,0)} \left( \tau ,\, k \right) $, defined in eq. (\ref{P-mn}), and required to evaluate the expression (\ref{YYL-par1}). We first solve the Einstein equations for the tensor metric perturbations, which, in terms of the canonical variables $\hat{h}_{c,\lambda}\equiv\frac{M_p a}{2}\hat{h}_\lambda $, read  
\begin{equation}
\Big(\partial_\tau^2+k^2-\frac{2}{\tau^2}\Big)h_{c,\lambda}\Big(\vec{k}\Big)=\Pi_{ij,\lambda}\Big(\hat{k}\Big)S_{ij}\Big( \tau, \vec{k}\Big) \equiv S_\lambda\Big( \tau,  \vec{k}\Big) \;, 
\label{eq-h}
\end{equation}
where the source $S_{ij}=-\frac{a^3}{M_p}\left(E_i E_j +B_i B_j\right)$ is due to the gauge mode $A_+$ amplified by the axion inflaton. This equation can be solved via the Green function method 
\begin{equation}
\hat{h}_{c,\lambda}=\hat{h}_{c,\lambda,vac}+\int^\tau_{-\infty}d\tau' G_k(\tau,\tau')S_\lambda\Big(\tau ', \vec{k}\Big) \;, 
\label{hc-formal} 
\end{equation}
where $\hat{h}_{c,\lambda,vac}$ is the vacuum mode (namely, the homogeneous solution of (\ref{eq-h})), and where the Green function associated with (\ref{eq-h}) is 
\begin{equation}
G_k \left( \tau ,\, \tau' \right) = \frac{1}{k^3 \tau \tau'} \left\{ \left( 1+k^2\tau\tau' \right) \sin \left[ k \left( \tau-\tau' \right) \right] + k \left( \tau'-\tau \right) \cos \left[ k \left( \tau-\tau' \right) \right] \right\} \,. 
\end{equation}

From this point on we disregard  the vacuum solution $\hat{h}_{c,\lambda,vac}$ since it leads to a vanishing $\langle R {\tilde R} \rangle$ (given that it is left/right symmetric) and since it is uncorrelated with the sourced modes. The two quantities $P_\lambda^{(2,1)}$ and $P_\lambda^{(1,0)}$ are then expressed as time derivatives of products of the integrals of two Green functions times a source correlator $\left\langle S_\lambda S_\lambda \right\rangle$. We evaluate this as done for the GW power spectrum in  \cite{Sorbo:2011rz,Barnaby:2011vw}, and obtain 
\begin{eqnarray}
P_\lambda^{(m,n)} \left( \tau ,\, k \right)  &=&  - \, \frac{H^4 }{M_p^4} \, 
\frac{k^3}{2 \pi^2} \, \frac{e^{4\pi\xi}}{\xi} \, \int \frac{d^3p_*}{(2\pi)^3}\sqrt{|\vec{p}_*| |{\hat k}-\vec{p}_*|} \,   f^{(m)} \left[ x ,\, \vec{p}_* ,\, \xi \right] \, f^{(n)} \left[ x ,\, \vec{p}_* ,\, \xi \right] \nonumber\\ 
&& \times  \frac{(1 - \lambda \lambda' \cos \theta)^2 \Big(1-p_*\cos\theta - \lambda \lambda' \sqrt{1-2p_*\cos\theta+p^2_*}\Big)^2}{16(1-2p_*\cos\theta+p^2_*)} \Big\vert_{x = - k \tau,\, \lambda'=+1}  \;, \nonumber\\ 
\label{P-mn-partial}
\end{eqnarray} 
where 
\begin{eqnarray} 
f^{(n)} \left[ x ,\, \vec{p}_* ,\, \xi \right] &\equiv&  \int_x^\infty d z \, \sqrt{z} \, \frac{d^n}{d x^n} \left[ \left( 1 + x z \right) \sin \left(z-x \right)+ \left( x - z \right) \, \cos \left( x - z \right) \right] \nonumber\\ 
&& \quad\quad \quad\quad \times 
\left[ \frac{2\xi}{z}+\sqrt{|\vec{p}_*| |{\hat k}-\vec{p}_*|} \right] \, 
e^{-2\sqrt{2\xi \, z} \left[ \sqrt{|\vec{p}_*|}+\sqrt{|{\hat k}-\vec{p}_*|} \right]} \,. 
\label{fn}
\end{eqnarray} 
We recall that $P_\lambda^{(0,0)}$ is the power spectrum of the helicity $\lambda$, and the result just reported  is analogous to the expression (3.40) of \cite{Barnaby:2011vw} for the power spectrum, with the only differences that here  we take the $m-th$ and $n-th$ derivatives on the two mode functions. The variable $x$ in the above expressions is the rescaled external time $x = - k \tau$, while the variable $z$ is the rescaled integration time $z = - k \tau'$ from eq. (\ref{hc-formal}). The variable $\vec{p_*}$ is a dimensionless momentum (obtained by dividing the internal momentum in the convolution for ${\cal S}_{ij}$ by the external momentum $k$). We have inserted $\lambda'$ to generalize this expression to the case in which the amplified gauge mode is $A_{\lambda'}$ (therefore, in the present work, $\lambda' = + 1$). 

Using (\ref{P-mn-partial}), eq. (\ref{YYL-par1}) rewrites 
\begin{eqnarray}
\!\!\!\!\!\!\!\! 
\frac{Y_L}{Y_{L,{\rm needed}}} & = & - 
8.8 \cdot 10^5 \,  \left( \frac{H}{M_p} \right)^{3/2} \, 
 \frac{1}{a^3 H^3} \nonumber\\ 
&&\times \, \int_{\tau_{\rm in}}^{\tau_{\rm end}} d \tau \int^{a \left( \tau \right) H}_{a_{in}H} \frac{d k}{k} \, \sum_{\lambda = \pm} 
 \lambda \, \frac{k^4 H^4}{2\pi^2 \, M_p^4}\frac{e^{4\pi\xi}}{\xi}  \,\int \frac{d^3p_*}{(2\pi)^3}\sqrt{|\vec{p}_*| |{\hat k}-\vec{p}_*|} \nonumber\\
 &&\times  \frac{(1 - \lambda \lambda' \cos \theta)^2 \Big(1-p_*\cos\theta - \lambda \lambda' \sqrt{1-2p_*\cos\theta+p^2_*}\Big)^2}{16(1-2p_*\cos\theta+p^2_*)}  \nonumber\\
&&\times\left(f^{(2)} \left[ x ,\, \vec{p}_* ,\, \xi \right]f^{(1)} \left[ x ,\, \vec{p}_* ,\, \xi \right]-f^{(1)} \left[ x ,\, \vec{p}_* ,\, \xi \right]f^{(0)} \left[ x ,\, \vec{p}_* ,\, \xi \right]\right) \;. 
\label{klimit}
\end{eqnarray}

The limits on the $k-$integration are due to the fact that only modes that have left the horizon during inflation (from the start of inflation to any given moment during inflation) contribute to the integral (this is the regime in which the $A_+$ mode is amplified, and  described by eq. (\ref{Apsol})). We elaborate on this point and justify this choice for the upper limit of integration in Appendix \ref{app:upperlimit}. It is convenient to change integration variables from $\left\{ \tau ,\, k \right\}$ to $\left\{ a = - \frac{1}{H \tau} ,\, x = - k \tau \right\}$. This gives 
\begin{eqnarray}
\frac{Y_L}{Y_{L,{\rm needed}}} & = & - 
\frac{8.8 \cdot 10^5}{2 \pi^2 a^3} \,  \left( \frac{H_{\rm end}}{M_p} \right)^{11/2} \, 
 \int_{0}^{a_{\rm end}} d a \, a^2  \, \left( \frac{H}{H_{\rm end}} \right)^7 
\, \frac{e^{4\pi\xi}}{\xi}\int^{1}_{0} \frac{d x}{x} \; 
 x^4\,\int \frac{d^3p_*}{(2\pi)^3}\sqrt{|\vec{p}_*| |{\hat k}-\vec{p}_*|} \nonumber\\
 &&\times \sum_{\lambda=\pm} \lambda \, \frac{(1 - \lambda \lambda' \cos \theta)^2 \Big(1-p_*\cos\theta - \lambda \lambda' \sqrt{1-2p_*\cos\theta+p^2_*}\Big)^2}{16(1-2p_*\cos\theta+p^2_*)}  \nonumber\\
&&\times\left(f^{(2)} \left[ x ,\, \vec{p}_* ,\, \xi \right]f^{(1)} \left[ x ,\, \vec{p}_* ,\, \xi \right]-f^{(1)} \left[ x ,\, \vec{p}_* ,\, \xi \right]f^{(0)} \left[ x ,\, \vec{p}_* ,\, \xi \right]\right) \,, 
\end{eqnarray}
where we have set $a_{\rm in} = 0$ at the start of inflation (this is irrelevant as the production is dominated by the latest time of inflation), and where the suffix ${\rm end}$ denotes the end of inflation. In the following, we normalize to one the scale factor at the end of inflation, $a_{\rm end} = 1$. Next, we define ${\tilde \lambda} = \lambda \, \lambda'$ and we change the sum over $\lambda$ into a sum over ${\tilde \lambda}$. We then perform a trivial angular integral in the $d^3 p_*$ integration, and we denote by $\theta$ the angle between ${\hat k}$ and $\vec{p}_*$. We arrive to 
\begin{eqnarray}
\!\!\!\!\!\!\!\!  \!\!\!\!\!\!\!\! 
\frac{Y_L}{Y_{L,{\rm needed}}} & = & - 
8.8 \cdot 10^5 \,  \left( \frac{H_{\rm end}}{M_p} \right)^{11/2} \, \int_{0}^{1} d a \, a^2  \, \left( \frac{H}{H_{\rm end}} \right)^7 \;  \frac{e^{4\pi\xi}}{\xi} \; G \left( \lambda' \right) \;, 
\label{YYL-app}
\end{eqnarray}
where we have defined 
\begin{eqnarray}
G \left( \lambda' \right) &\equiv&  \frac{\lambda'}{2 \pi^2} \, \int^{1}_{0} d x  \, x^3 \, 
\int_0^{\infty} \frac{d p_* \, p_*^2}{(2\pi)^2} \int_{-1}^{+1} d \cos \theta \, p_*^{1/2} \, 
\left( 1 - 2 p_* \cos \theta + p_*^2 \right)^{1/4}  \nonumber\\
&& \!\!\!\!\!\!\!\!  \!\!\!\!\!\!\!\! 
\times\left(f^{(2)} \left[ x ,\, \vec{p}_* ,\, \xi \right]f^{(1)} \left[ x ,\, \vec{p}_* ,\, \xi \right]-f^{(1)} \left[ x ,\, \vec{p}_* ,\, \xi \right]f^{(0)} \left[ x ,\, \vec{p}_* ,\, \xi \right]\right)\nonumber\\ 
&& \!\!\!\!\!\!\!\!  \!\!\!\!\!\!\!\! 
\times \sum_{\tilde \lambda = -1}^{1} \; {\tilde \lambda} \; \frac{(1 - {\tilde \lambda} \, \cos \theta)^2 \Big(1-p_*\cos\theta - {\tilde \lambda} \, \sqrt{1-2p_*\cos\theta+p^2_*}\Big)^2}{16(1-2p_*\cos\theta+p^2_*)} 
\label{Gl}
\end{eqnarray}

We are interested in the $\xi \gg 1$ limit of this expression. One can verify numerically that the momentum integral is dominated at $p_* = {\rm O } \left( 1 \right)$.  Therefore, due to the exponential term, the integral is dominated by the region $z \ll 1$ of the integrand. Using this, we can obtain an accurate analytic expression for $f^{(n)}$. Specifically, starting from eq. (\ref{fn}),  we first take the $x$ derivative, and then Taylor expand the term that multiplies the exponential, obtaining an expression of the form 
\begin{equation}
f^{(n)} \simeq \int_x^\infty \frac{d z}{\sqrt{z}} \left[ c_0 + c_1 z + c_2 z^2 + c_3 z^3 + \dots \right] e^{-2\sqrt{2\xi \, z} \left[ \sqrt{|\vec{p}_*|}+\sqrt{|{\hat k}-\vec{p}_*|} \right]} \,, 
\label{fz-Taylor} 
\end{equation} 
where the $c_n$ coefficients are $x-$dependent.  Each term in this expansion can be integrated analytically, using 
\begin{equation} 
\int_x^\infty \frac{d z}{\sqrt{z}} z^n \, {\rm e}^{-a \sqrt{z}} = \frac{2}{a^{1+2 n}} \, \Gamma \left( 1 + 2 n ,\, a \sqrt{x} \right) \,. 
\end{equation} 
As remarked, the integral (\ref{fn}) is dominated by the region at $z \ll 1$. Therefore, the various terms $c_n z^n$ in the Taylor expansion become progressively less relevant as $n$ increases (we verified numerically that the terms with $n > 6$ can be neglected). 

With this analytic expressions, the function $G \left( \lambda' \right)$ contains a three dimensional integral that we evaluated numerically for several values of $\xi$. The result can be well fitted by 
\begin{equation}
G \left( \lambda' \right) \simeq - \lambda' \; \frac{5.75 \cdot 10^{-8}}{\xi^{5.52}} \;\;\;,\;\;\; 3 \leq \xi \leq 7 \;, 
\end{equation} 
in the regime of $\xi$ that is relevant for our computation (see the main text). Inserting this expression into 
(\ref{YYL-app}) we obtain eq. (\ref{YYL-par2}) of the main text (where we set $\lambda' = +1$, as this is the case considered in this work).

\section{Upper limit of the physical momentum integral}
\label{app:upperlimit}

In this appendix we justify our choice of including in the integral for the lepton asymmetry only gravitational waves that have left the horizon during inflation. This is done by setting  $k \leq a(\tau) H$ in equation (\ref{klimit}), or, equivalently,  $x \leq 1$ in eq. (\ref{Gl}).  In the integrand of eq. (\ref{Gl}), the growing monomial $x^3$ is more than compensated by the exponential decrease of the functions $f^{(n)} \left[ x ,\, \vec{p}_* ,\, \xi \right] $ in the $x \geq 1$ regime. Mathematically, the exponential decrease is due to the factor  $e^{-2\sqrt{2\xi \, z} \left[ \sqrt{|\vec{p}_*|}+\sqrt{|{\hat k}-\vec{p}_*|} \right]}$ present in eq. (\ref{fz-Taylor}), After carrying out the $z$ integration of eq. (\ref{fz-Taylor}), the upper $z=\infty$ extremum gives no contribution, while the lower $z=x$ extremum gives a contribution proportional to $e^{-2\sqrt{2\xi \, x} \left[ \sqrt{|\vec{p}_*|}+\sqrt{|{\hat k}-\vec{p}_*|} \right]}$. We note that this contribution is progressively more suppressed at increasing $\xi$, and it is extremely effective for all values of $\xi$ needed to produce a non-negligible asymmetry. 
 
The physical origin of the suppression resides in the gauge field amplification (originating from the $\phi F {\tilde F}$ mechanism) that gives rise to the amplitude (\ref{Apsol}) for the $A_+$ mode.  As seen from this equation, only super-horizon modes of the vector field are produced (mathematically, the negative term in the exponent becomes greater than the positive term for sub-horizon modes). These super-horizon vector modes source at an appreciable level only metric tensor perturbations that are super-horizon (namely, no high momentum gravitational mode is produced by the ``fusion'' of two low momentum vector modes in the sourcing process $A + A \rightarrow h$). The suppression in the production of sub-horizon gravitational waves is mathematically encoded by the exponential suppression of the $f^{(n)}$ functions at $x > 1$ discussed in the previous paragraph.  

To appreciate this, and to quantify the residual effect of the sub-horizon modes, in Figure $2$ we show the evolution of the ratio 
\begin{equation}
R \left( x_{\rm max} \right) \equiv \frac{G \left( \lambda' \right) \;\; {\rm with \; upper \; extremum \; of \; integration } \; x = x_{\rm max}}{G \left( \lambda' \right) \;\; {\rm with \; upper \; extremum \; of \; integration } \; x = 1} \,. 
\label{R-xmax}
\end{equation} 
This quantity (normalized to one for $x_{\rm max}=1$), describes how our result for $G \left( \lambda' \right)$ is sensitive to our choice of the upper extremum of integration $x$ in eq. (\ref{Gl}).  
\begin{figure}[ht!]
\centerline{
\includegraphics[width=0.5\textwidth,angle=0]{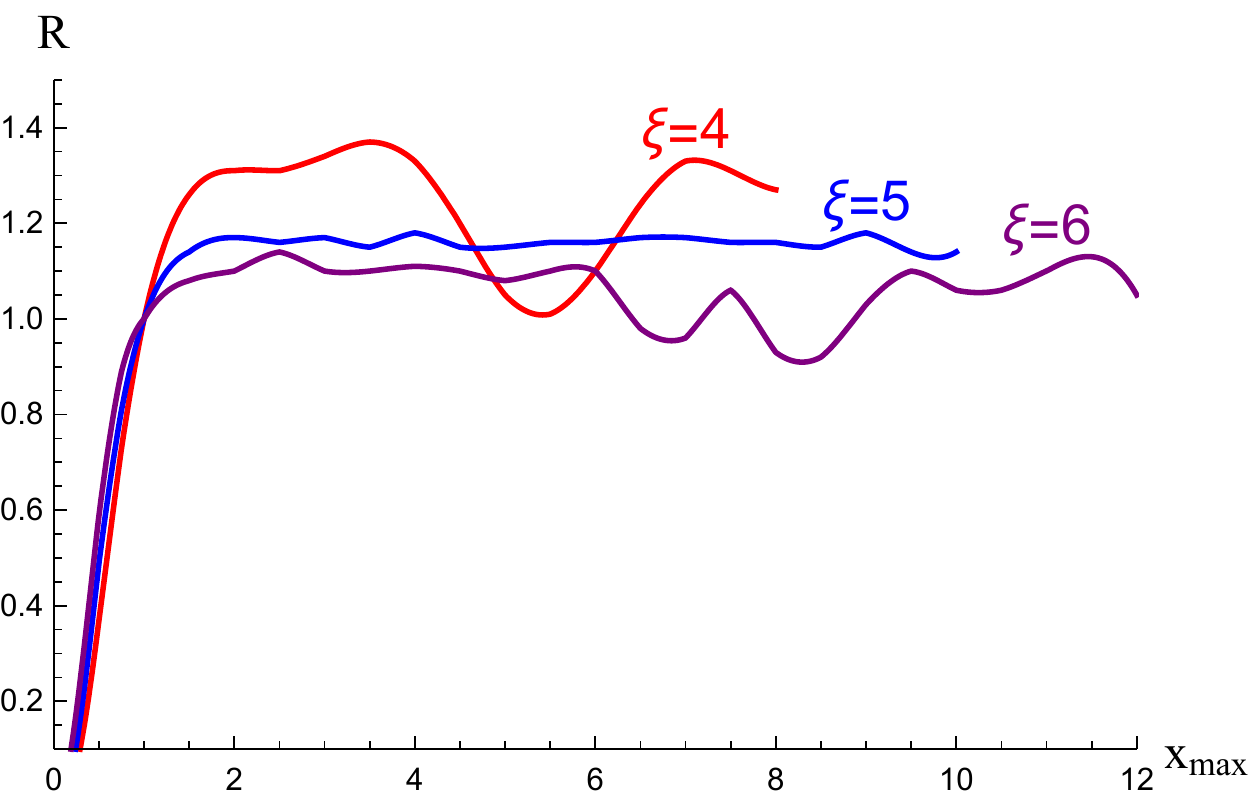}
}
\caption{Dependence of the lepton asymmetry on the choice of the UV cut-off $x_{\rm max}$ in eq. (\ref{Gl}). The ratio $R \left( x_{\rm max} \right)$ shown in this figure is defined in eq. (\ref{R-xmax}).}
\label{fig:graph}
\end{figure}

Extrapolating the result shown in the figure in the $x_{\rm max} \ll 1$ regime shows that  modes that are much larger than the horizon at the end of inflation ($x \ll 1$) give a negligible contribution to the asymmetry. Therefore, to obtain an accurate estimate of the asymmetry, one indeed needs to integrate at least up to $x_{\rm max} \simeq 1$. This is what we have done in the computations of Appendix \ref{app:correl}. The figure also proves that the inclusion of modes with $x > 1$ would increase the result obtained in Appendix \ref{app:correl} by $\sim 30\%$ at most in the case of $\xi = 4$ and by $\sim 10\%$ at most in the case of $\xi = 6$ (as we discussed at the end of the first paragraph of this appendix, the contribution of the $x>1$ modes is progressively reduced as $\xi$ increases). We see from Table \ref{table:results} that, even if $\xi \ga 6 $ at the end of inflation, the asymmetry obtained by fixing $x_{\rm max} =1$ is still $\sim 3$ orders of magnitude smaller than the required asymmetry. Even a order one increase (which is much more than what is shown in the figure) therefore would not change any of our results. 

We stress that this discussion is strongly linked with the form of the gauge field solution (\ref{Apsol}) used in this paper. This solution, originally obtained in \cite{Anber:2006xt}, is accurate in the $\left(8\xi\right)^{-1} \lesssim x \lesssim 2\xi$ interval \cite{Anber:2006xt,Barnaby:2011vw}. The lines shown in Figure \ref{fig:graph} are limited to this region, and they clearly show that the contribution of modes in the $1 \leq x \leq 2 \xi$ region is subdominant to that in the $0 \leq x \leq 1$ one. Extending the computation to $x > 2 \xi$ requires the use of more appropriate solutions. In particular, it requires a UV regularization of the gauge fields produced by the abelian $\phi F {\tilde F}$ mechanism. To see this more explicitly, we recall that the gauge modes obey the equation 
\begin{equation}
\left( \partial_x^2 + 1 \mp \frac{2 \xi}{x} \right) A_\pm \left( x \right) = 0 \;. 
\label{eqA}
\end{equation}
Gauge field amplification in the abelian $\phi F {\tilde F}$ mechanism is due to the last term in this equation. This term is shut-off in the UV regime $x \gg 2 \xi$, where the equation (\ref{eqA}) reduces to the vacuum equation $\left( \partial_x^2 + 1 \right) A_\pm \left( x \right) = 0$, which is solved by the vacuum modes $A_\pm = \frac{{\rm e}^{i x}}{\sqrt{2 k}} = \frac{{\rm e}^{-i k \tau}}{\sqrt{2 k}}$. The net contribution to the asymmetry from the vacuum modes vanishes (as the vacuum $\pm$ chiralities are produced in equal amount). Moreover any physical effect associated with these modes needs to be regularized. For instance, their physical energy density satisfies 
\begin{equation}
\left\langle \frac{E^2 + B^2}{2} \right\rangle = \frac{1}{4 \pi^2 a^4} \sum_{\lambda=\pm} \, \int d k \, k^2 \left[ \vert \partial_\tau A_\lambda \vert^2 + k^2 \vert A_\lambda \vert^2 \right] =  \frac{1}{2 \pi^2 a^4} \int d k \, k^3 \;, 
\end{equation} 
which (as it is well known) diverges in the UV. On the contrary, the approximate solution (\ref{Apsol}) accounts for the physical gauge amplification at $x < 1$, without having the unphysical UV divergence associated to the vacuum modes.~\footnote{Although not often explicitly  stated, the regularization provided by the solution (\ref{Apsol}) is implicitly assumed in the vast majority of the works that study the physical effects of gauge field amplification by the abelian $\phi F {\tilde F}$ mechanism (such as scalar non-gaussianity and running of the spectral tilt, gravitational waves, primordial black holes). None of these works computes the contribution from unregularized vacuum modes in the deep UV regime (which, we stress, is a contribution unrelated to the gauge field amplification from this mechanism).  With this understanding, we also employ the solution  (\ref{Apsol}) in this work. } 

In summary, figure \ref{fig:graph} shows that integrating up to $x_{\rm max} =1$ provides a very good estimate of the asymmetry due to gauge field amplification in the abelian $\phi F {\tilde F}$ mechanism. Greater momentum modes are not amplified by this mechanism \cite{Anber:2006xt,Barnaby:2011vw}, and they reduce to the vacuum modes, which do not contribute to the asymmetry (after summing over the two chiralities), and whose effect needs anyhow to be regulated away in the UV.

\end{document}